\documentclass[aps,nofootinbib,superscriptaddress, showpacs,preprintnumbers,nofootinbibt,twocolumn]{revtex4-2}
\usepackage{amsmath}
\usepackage{epsfig}
\usepackage{multirow}
\usepackage{adjustbox}
\usepackage{eurosym}
\usepackage{dcolumn}
\usepackage{bm}
\usepackage{enumerate}
\usepackage{float}
\usepackage{epstopdf}
\usepackage{amsmath}
\usepackage{bm}
\usepackage{natbib}
\usepackage{amsfonts}
\usepackage{amssymb}
\usepackage{graphicx}
\usepackage{alphalph,mathtools}
\usepackage{etoolbox}
\usepackage{color}
\usepackage{booktabs}
\usepackage{footnote}
\usepackage{makecell,tabularx}
\usepackage{hyperref}
\usepackage{subfigure, rotating, bm, array}
\hypersetup{colorlinks,citecolor=blue}
\hypersetup{colorlinks=true,linkcolor=red,filecolor=magenta,
    urlcolor=blue}

\def\be{\begin{equation}}
\def\ee{\end{equation}}
\def\bea{\begin{eqnarray}}
\def\eea{\end{eqnarray}}

\begin{document}
\title{Exploring Tidal Force Effects and Shadow Constraints for Schwarzschild-like Black Hole in Starobinsky-Bel-Robinson Gravity} 
\author{Dhruv Arora}
\email{arora09dhruv@gmail.com}
\affiliation{Pacif Institute of Cosmology and Selfology (PICS), Sagara, Sambalpur 768224, Odisha, India}
\author{Niyaz Uddin Molla}
\email{niyazuddin182@gmail.com} 
\affiliation{Department of
Mathematics, Indian Institute of Engineering Science and
Technology, Shibpur, Howrah-711 103, India,}
\author{Himanshu Chaudhary}
\email{himanshuch1729@gmail.com} 
\affiliation{Department of Applied Mathematics, Delhi Technological University, Delhi-110042, India,}
\affiliation{Pacif Institute of Cosmology and Selfology (PICS), Sagara, Sambalpur 768224, Odisha, India}
\affiliation{Department of Mathematics, Shyamlal College, University of Delhi, Delhi-110032, India,}
\author{Ujjal Debnath}
\email{ujjaldebnath@gmail.com} \affiliation{Department of
Mathematics, Indian Institute of Engineering Science and
Technology, Shibpur, Howrah-711 103, India,}
\author{Farruh Atamurotov}
\email{atamurotov@yahoo.com} 
\affiliation{New Uzbekistan University, Mustaqillik ave. 54, 100007 Tashkent, Uzbekistan}
\affiliation{Central Asian University, Milliy Bog' Street 264, Tashkent 111221, Uzbekistan}
\affiliation{Faculty of Computer Engineering, Tashkent University of Applied Sciences, Gavhar Str. 1, Tashkent 100149, Uzbekistan}
\affiliation{Institute of Theoretical Physics, National University of Uzbekistan, Tashkent 100174, Uzbekistan}
\author{G.Mustafa}
\email{gmustafa3828@gmail.com}
\affiliation{Department of Physics,
Zhejiang Normal University, Jinhua 321004, People’s Republic of China}
\affiliation{Zhejiang Institute of
Photoelectronics and Zhejiang Institute for Advanced Light Source,
Zhejiang Normal University, Jinhua, Zhejiang 321004, China}

\begin{abstract}
The current manuscript deals with the tidal force effects, geodesic deviation, and shadow constraints of the  Schwarzschild-like black hole theorised in Starobinsky-Bel-Robinson gravity exhibiting M-theory compactification. In the current analysis, we explore the radial and angular tidal force effects on a radially in-falling particle by the central black hole, which is located in this spacetime. We also numerically solve the geodesic deviation equation and study the variation of the geodesic separation vector with the radial coordinate for two nearby geodesics using suitable initial conditions. All the obtained results are tested for Sag A* and M87* by constraining the value of the stringy gravity parameter $\beta$ using the shadow data from the event horizon telescope observations. All the results are compared with Schwarzschild black hole spacetime. In our study, we found that both the radial and angular tidal forces experienced by a particle switch their initial behaviour and turn compressive and stretching, respectively, before reaching the event horizon. The geodesic deviation shows an oscillating trend as well for the chosen initial condition. For the constrained value of $\beta$, we see that the spacetime geometry generated by Sag A* and M87* is effectively same for both Schwarzschild and Starobinsky-Bel-Robinson black hole. Furthermore, we also calculated the angular diameter of the shadow in Starobinsky-Bel-Robinson black hole and compared with the Schwarzschild black hole. It is observed that the angular diameter of shadow for M87* and Sgr A* in Starobinsky-Bel-Robinson black hole is smaller than the Schwarzschild black hole. The calculated results satisfy the event horizon telescope observational constraints. 
\end{abstract}

\maketitle

\section{INTRODUCTION}
Black holes are one of the most fascinating components in our universe, representing a fundamental prediction within Einstein's theory of general relativity. The scientific community and researchers have been intrigued by black holes for numerous years, driven by various reasons. Their captivating nature makes them curious and exotic entities within the universe. A widely accepted theory is that black holes are formed through the gravitational collapse of massive stars \cite{chandrasekhar1998mathematical}. Another possible alternative to the astrophysical origin of black holes are primordial black holes, formed due to over-densities seeded during inflation or by topological defects during radiation dominated era of the universe. \cite{carr2010new,sasaki2018primordial,niemeyer1999dynamics,kristiano2022ruling}. The existence of black holes has been strongly supported by several observational characteristics. Notably, measurements of black hole spin in X-ray binaries, the detection of gravitational wave signals from binary black hole mergers by LIGO \cite{LIGOScientific:2016aoc,LIGOScientific:2020iuh,LIGOScientific:2020stg}, and the remarkable image of a black hole at the core of the galaxy M87 captured by the Event Horizon Telescope (EHT) collaboration \cite{2} have provided compelling evidence. Additionally, the discovery of a wide star-black hole binary system through radial velocity measurements \cite{Liu:2019lfc} has further contributed to our comprehensive understanding of black holes. These achievements collectively reinforce the evidence for the existence and properties of black holes. These enigmatic objects have captured considerable attention across the fields of astrophysics, astronomy and high-energy physics. Through studying black holes, researchers have made remarkable discoveries in areas such as thermodynamics, quantum effects, and gravitational interactions within curved spacetime. The physical properties of black holes remain a fascinating subject for the scientific community, even after the first discovery of a black hole as a solution to Einstein's equations more than a hundred years ago in General Relativity (GR). This initial discovery gave rise to the well-known Schwarzschild solution \cite{Schwarzschild:1916uq}, characterized by a single parameter - the mass of the black hole.
Since then, multiple solutions to Einstein's equations have been unveiled, going beyond the Schwarzschild solution. These encompass the Kerr black hole, exhibiting rotation \cite{Kerr:1963ud}, the Reissner-Nordström black hole, demonstrating electrical charge \cite{reissner1916eigengravitation,nordstrom1918energy}, and the Kerr-Newman black hole, possessing both charge and rotation \cite{Newman:1965my}. Thus
in the context of GR,  black holes can be described using three fundamental parameters: mass, charge, and angular momentum. However, when it comes to astrophysical black holes, it is generally anticipated that they are characterized by only two of these parameters - mass and angular momentum. Despite this simplification, these black holes may exist within highly dynamic environments, featuring elements like accretion disks and electromagnetic fields \cite{frank2002accretion}. Furthermore, in modified theories of gravity, black holes are linked to slight deviations from their standard counterparts, leading to unique features in their behavior \cite{Berti:2015itd}. Investigating how these deviations can be observed in electromagnetic and gravitational wave observations is currently a highly active area of research in the scientific literature \cite{Barausse:2014tra,Berti:2019wnn,Ferreira:2017pth}. In the last decade, modified gravity models have been extensively explored to address various cosmological issues and investigate the properties of black holes. Numerous theories have been proposed, taking diverse approaches and methods. Among these, Einstein Gauss-Bonnet (EGB) gravity, which finds support from string theory and related dual models, including M-theory \cite{Wheeler:1985nh,ghosh2020generating,Ghosh:2020vpc}, has emerged as one of the most extensively studied theory. In the context of EGB gravity, black holes have been investigated with respect to the Gauss-Bonnet gravity parameter. This parameter plays a pivotal role in constructing and studying black holes within these gravity models \cite{Ghosh:2020ijh,singh20214d}. The thermodynamics and optical properties of these black holes have been carefully examined concerning the Gauss-Bonnet parameter\cite{belhaj2022optical,Belhaj:2022qmn}. Notably, it has been observed that variations in the Gauss-Bonnet parameter significantly impact the physical properties of black holes, with the shadow radius decreasing as the parameter increases \cite{Vagnozzi:2022moj}. These findings provide valuable insights into the interplay between gravity models and the intriguing behavior of black holes, furthering our understanding of the cosmos at both macroscopic and microscopic scales. Recently, a novel gravitational model called Starobinsky-Bel-Robinson (SBR) gravity has been introduced, incorporating a new parameter denoted by $\beta$ \cite{Ketov:2022lhx}. This four-dimensional gravity is inspired by M-theory, which, at lower energies, is described by eleven-dimensional supergravity. Specifically, there are suggestions that this modified gravity is inspired from M-theory compactified on a two-sphere factor represented by $S^3 \times S^4$ \cite{Ketov:2022lhx}. Various models have been proposed within this novel gravitational framework. Inflation scenarios involving the $\beta$  parameter have been studied, and associated cosmological observables have been computed \cite{ketov2022superstring}. These investigations have explored how observational data can impose constraints on the value of $\beta$.
Furthermore, Schwarzschild-type black holes in SBR gravity have been constructed, and their thermodynamic properties have been analyzed in terms of the parameter $\beta$. Relevant quantities, including entropy and pressure, have been computed, revealing that the presence of $\beta$  introduces corrections to these quantities \cite{delgado2022schwarzschild}. These explorations shed light on the influence of the $\beta$  parameter on the thermodynamic properties of black holes in the context of SBR gravity. In a recent study, Belhaj et al. \cite{belhaj2023deflection} investigated the deflection angle and shadows produced by black holes within the framework of Starobinsky-Bel-Robinson Gravity inspired from M-theory.\\\\
In the current paper, we focus on the analysis of tidal forces
in the black holes in  Starobinsky-Bel-Robinson modified Gravity from M-theory, recently obtained by Delgado and  Ketov \cite{delgado2022schwarzschild}. It is widely known that tidal forces within a Schwarzschild spacetime cause a body falling towards the event horizon to experience stretching in the radial direction while being compressed in the angular directions
\cite{d1992clarendon,itin2008mp}. The stretching and compression are induced by gravity's tidal effect, which is created by a difference in gravity's strength between two neighbouring places. This phenomenon is highly widespread in our cosmos and have sparked public scientific interest for most of the twentieth century. In the last decades, tidal forces
were analyzed for Kiselev black hole \cite{Shahzad:2017vwi}, Hayward black hole spacetime \cite{lima2020tidal}, null naked singularity spacetime \cite{madan2022tidal} and some
regular black holes \cite{sharif2018tidal}. The authors of \cite{gad2007geodesics} compared the tidal force effects in a stringy charged black hole to those in Schwarzschild and RN black holes. The authors of \cite{Crispino:2016pnv} studied tidal forces in Reissner-Nordström spacetime and discovered that the radial and angular tidal forces switches sign at the event horizon. Also, the tidal forces in presence of a cosmological constant were studied in \cite{vandeev2021tidal}.\\\\
The tidal forces in the Kerr Black Hole were
investigated in Refs.\cite{marck1983solution,tsoubelis1988inertial,Chicone:2003yv,Chicone:2004ic,Chicone:2004pv,LimaJunior:2020fhs}. 
Tidal forces play a significant role in astrophysics, leading to intriguing phenomena such as tidal disruption events (TDEs). In these events, a star may be disrupted by the tidal forces exerted by a black hole\cite{Kesden:2011ee} ,giving rise to bright flares of X-ray \cite{bade1996detection}, ultraviolet \cite{gezari2008uv}, and optical \cite{vanVelzen:2010jp} radiation. These TDEs offer valuable insights into the interactions between black holes and surrounding celestial bodies.\\\\
In this paper, we thoroughly analyze the tidal forces and geodesic deviation for a radially in-falling particle in the Schwarzschild-like black hole spacetime in the Starobinsky-Bel-Robinson Gravity. Furthermore, we also find the constraint of the black hole parameter $\beta$ with the black hole shadow by the supermassive black holes $M87^*$ and  $SgrA^*$ data. The shadow of a black hole is created by photons emitted from the central black hole and eventually reaching an observer. These photons follow unstable orbits, forming light rings around the black hole. When projected onto an observing screen, these light rings give rise to the shadow. Unstable orbits are critical because small perturbations can cause photons to either be captured by the central compact object or escape to infinity. The light rays composing the shadow pass very close to the event horizon and experience significant bending due to the strong gravitational lensing effect. Consequently, observations of the shadow provide a means to test the strong-field properties of gravity. The seminal work of Synge \cite{39} and Luminet \cite{Luminet:1979nyg} was instrumental in developing a formula for calculating the angular radius of photons. This formula, in turn, facilitated the characterization of the region surrounding a Schwarzschild black hole by elucidating the behavior of the light deflection angle as it diverges. Following this, Bardeen's\cite{bardeen1973rapidly} groundbreaking research in 1973 delved into the intricacies of the shadow cast by a spinning Kerr black hole, demonstrating how the black hole's rotation caused distortions in the shadow's shape. Within this context, the photon ring, encircling the black hole's shadow, emerges as a pivotal element that can be harnessed to determine the black hole's parameters and unveil gravitational features within the information field. Subsequently, there has been a surge of activity in both analytical/numerical investigations and observational studies related to the shadows cast by the different kind of black holes \cite{Vagnozzi:2022moj,Kumar:2020yem,Afrin2023aadada,pulicce2023constraints,Yan:2023pxj,ghorani2023probing,atamurotov2023weak,rec2,rec3,rec4}. Numerous studies have been conducted in this area, including those by the works of literature \cite{Perlick2022rev,mustafa2022shadows,pantig2022shadow,banerjee2022signatures,Molla:2022izk,atamurotov2022shadow,Atamurotov21b}.\\\\
The organization of this paper is as follows. In section \ref{II}, we reconsider the nature of the black holes in the SBR gravity. In Section \ref{III} we discuss the radial and angular tidal forces in SBR black hole spacetime using tetrad formalism. Section \ref{IV} presents the numerical solution for the geodesic deviation equation and compares the results with Schwarzschild spacetime. In section \ref{V}, we investigate the shadow behaviors and provide predictions for $\beta$ by the help of EHT data. We conclude our results in section \ref{VI}. Throughout the manuscript, we use as $h=1=c$.

\section{BLACK HOLES IN SBR GRAVITY}\label{II}
In this section, we discuss about the Schwarzschild like black hole solution in SBR gravity which is theorized to be embedded in M-theory existing in eleven dimensional spacetime. This solution is discussed by the authors in \cite{ketov2022starobinsky}. M-theory has a bosonic sector having a metric $g_{M N}$ and a tensor 3 -form $C_{M N P}$ coupled to M2-branes which are dual to M5-branes \cite{witten1997solutions}. Using the compactification process coupled with the existence of stringy fluxes, the corresponding 4-D gravity models may be produced \cite{ketov2022starobinsky,ketov2022superstring,
delgado2022schwarzschild}. The action considered is given by:

\begin{equation}
S_{S B R}=\frac{M_{p l}^{2}}{2} \int d^{4} x \sqrt{-g}\left(R+\frac{R^{2}}{6 m^{2}}-\frac{\beta}{32 M_{pl}^{6}}\left(P_{4}^{2}-E_{4}^{2}\right)\right).  
\label{1}
\end{equation}
where $g$ is the metric determinant and $R$ is the Ricci curvature scalar. $m$ is a free mass parameter which could have various interpretations depending on the underlying theory. $\beta$ is a positive dimensionless coupling whose value is determined by the compactification of M-theory, and could be fixed by studying the optical behaviors of black hole. $P_{4}^{2}$ and $E_{4}^{2}$ are the Pontryagin and the Euler topological densities which are related to the Bel-Robinson tensor $T_{\mu \nu \lambda \rho}$ in four dimensions by means of the relation \cite{ketov2022starobinsky,ketov2022superstring,
delgado2022schwarzschild}

\begin{equation}
T^{\mu \nu \delta \rho} T_{\mu \nu \delta \rho}=\frac{1}{4}\left(P_{4}^{2}-E_{4}^{2}\right).  
\label{2}
\end{equation}

The Bel-Robinson tensor is defined as:

\begin{equation}
T^{\mu \nu \delta \sigma}=R^{\mu \rho \gamma \delta} R_{\rho \gamma}^{\nu \sigma}+R^{\mu \rho \gamma \sigma} R_{\rho \gamma}^{\nu \delta}-\frac{1}{2} g^{\mu \nu} R^{\rho \gamma \alpha \delta} R_{\rho \gamma \alpha}^{\sigma}.
\label{3}
\end{equation}
It can be seen from Eq.(\ref{1}) that the gravity action depends on two parameters ($m,\beta$). This gives rise to plethora of applications such as studying Hawking radiation \cite{ketov1991curvature}, entropy \cite{elgood2021first} and inflation \cite{ivanov2022analytic}. The black hole solution in SBR gravity depends on the value of $\beta$ corrected to first order perturbations. The line element of this non-rotating solution has been found to be \cite{delgado2022schwarzschild}

\begin{equation}
d s^{2}=-f(r) d t^{2}+\frac{1}{f(r)} d r^{2}+r^{2} d \Omega^{2}.
\label{4}
\end{equation}
where the metric component $f(r)$ is given by

\begin{equation}
f(r)=1-\frac{r_{s}}{r}+\beta\left(\frac{4 \sqrt{2} \pi  G r_{s} }{r^{3}}\right)^{3}\left(\frac{108 r- 97 r_{s}}{5 r}\right).
\label{5}
\end{equation}
Here, $r_{s} =2GM$ is Schwarzschild radius, where $M$ represents the mass of a black hole. Taking $\beta= 0$, we recover the Schwarzschild spacetime metric. 

\section{TIDAL FORCE EFFECTS IN SBR BLACK HOLES}\label{III}
In this section, we will delve into the complexities of the equations involving tidal forces in our spacetime. To investigate the equation for the distance between two infinitesimally close and free falling particles, we use the following equation for the spacelike components of the geodesic deviation vector $\eta^{\mu}$
\begin{eqnarray}
    \frac{D^{2} \eta^{\mu}}{D\tau^{2}} = R^{\mu}_{\alpha \beta \gamma} v^{\alpha}  v^{\beta} \eta^{\gamma},
    \label{6}
\end{eqnarray}
where $R^{\mu}_{\alpha \beta \gamma}$ is the Riemann curvature tensor and $v^{\mu}$ is the unit tangent vector to the geodesic. It is known that a test body moving in space solely under the influence of gravity follows a geodesic \cite{Falcon-Gomez:2022wze}. Now, due to the difference in curvature at every point on the geodesic and the fact that each point on the test body tends to follows a unique geodesic, the body experiences a difference in acceleration, leading to a stretching and squeezing effect known as tidal forces. This is the reason that the Riemann curvature tensor is used to study the tidal interactions due to gravity.\\\\
To calculate the tidal forces in the falling body's frame, we make use of the tetrad formalism \cite{Mitsou:2019nhj}. Tetrads are geometric objects that form a set of local coordinate bases, i.e. a locally defined set of four linearly independent vector fields known as tetrads or vierbien. At each point on a geodesic, there is a tetrad frame that forms a local inertial reference frame, where the laws of special relativity apply. Since, Lorentz transformations can connect an infinite number of orthonormal bases at a specific point, they cannot give enough information about the connection on the metric. To resolve this, we use the transformation from the orthonormal basis to the coordinate basis:
\begin{eqnarray}
    \overrightarrow{e}_{\mu} = \hat{e}^{\hat{i}}_\mu  \overrightarrow{e}_{\hat{i}},
    \label{7}
\end{eqnarray}
where $\overrightarrow{e}_{\mu}$ represents the coordinate basis, $\overrightarrow{e}_{\hat{i}}$ represents the orthonormal basis and $\hat{e}^{\hat{i}}_\mu $ are the tetrad components. The metric tensor components in the tetrad basis are given as \cite{Mitsou:2019nhj}:
\begin{eqnarray}
    g_{\mu\nu} = \eta_{\hat{i}\hat{j}} \hat{e}^{\hat{i}}_\mu \hat{e}^{\hat{\hat{j}}}_\nu. 
    \label{8}
\end{eqnarray}
This is the key equation that enables us to use orthonormal basis in curved spacetime. The tetrad components for freely falling frames can be obtained from Eq. (\ref{7}) and Eq. (\ref{8}).
\begin{equation}
    \hat{e}^{\mu}_{\hat{0} } = \bigg\{\frac{E}{f(r)} , -\sqrt{E^{2}-f(r)}, 0, 0\bigg\} ,
    \label{9}
\end{equation}
\begin{equation}
    \hat{e}^{\mu}_{\hat{1} } = \bigg\{\frac{-\sqrt{E^{2}-f(r)}}{f(r)}, E, 0, 0\bigg\},
    \label{10}
\end{equation}
\begin{equation}
    \hat{e}^{\mu}_{\hat{2} } = \bigg\{0, 0, \frac{1}{r}, 0\bigg\}, 
    \label{11}
\end{equation}
\begin{equation}
    \hat{e}^{\mu}_{\hat{3} } = \bigg\{0 ,0, 0, \frac{1}{r\sin\theta}\bigg\}, 
    \label{12}
\end{equation} 
where the tetrad components follow the following rule:
\begin{equation}
     \hat{e}^{\mu}_{\hat{i} }  \hat{e}^{\hat{j}}_{\mu}  = \delta^{\hat{j}}_{\hat{i}}.
     \label{13}
\end{equation}
We notice that the component $\hat{e}^{\mu}_{\hat{0} } = v^{\mu}$ is the tangent vector to the geodesic i.e. the four velocity of the particle. Also, the geodesic deviation vector follows the transformation rule from global coordinates to local orthonormal coordinates:
\begin{equation}
    \eta^{\mu} = \hat{e}^{\mu}_{\hat{i} } \eta^{\hat{i}}.
    \label{14}
\end{equation}

\begin{figure}[h]
    \centering
    \includegraphics[width = \linewidth]{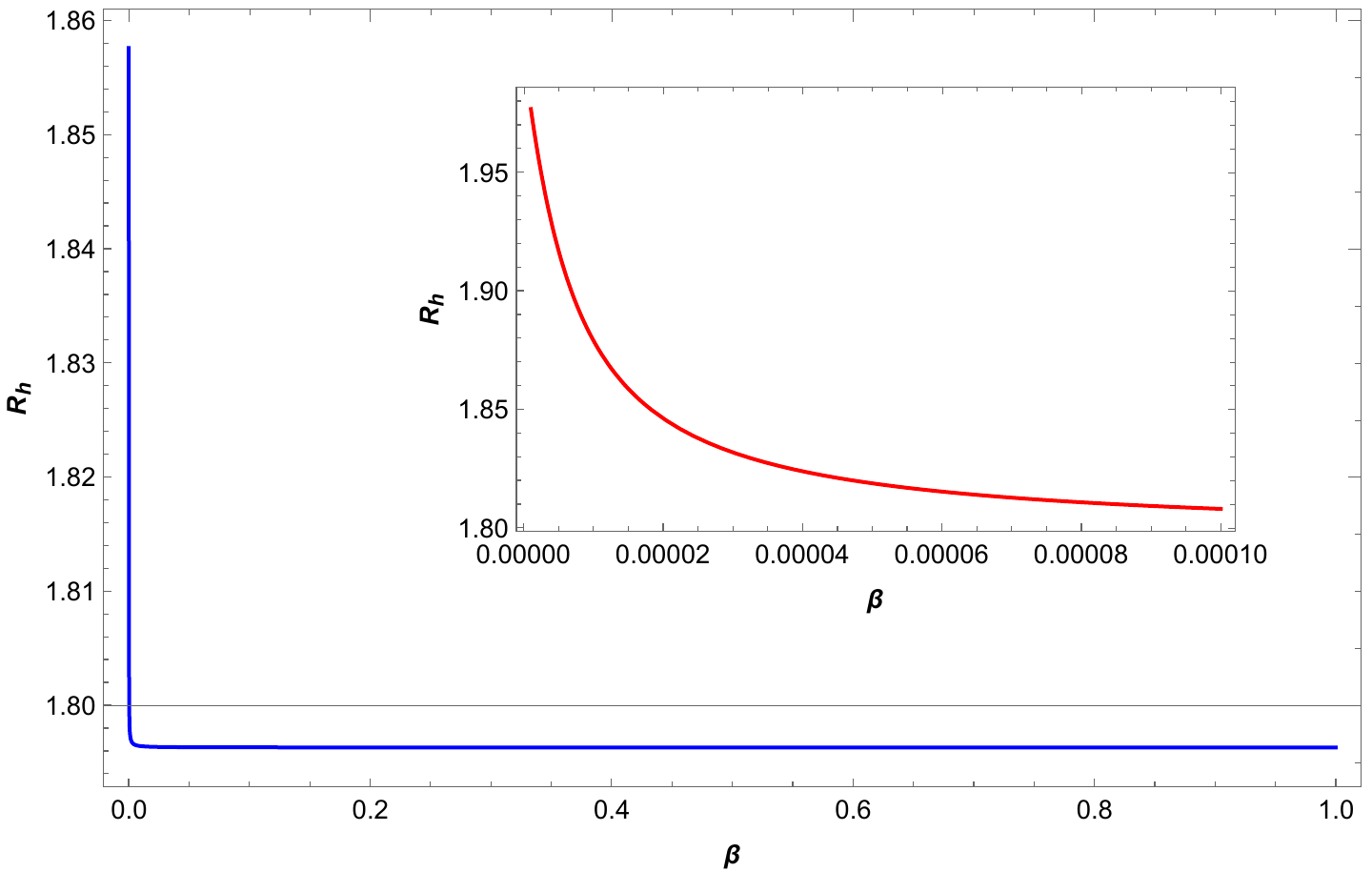}
    \caption{Event horizon location in SBR black holes for different values of stringy parameter $\beta$. We take M=1 for computation. }
    \label{fig:1}
\end{figure}
It is evident from Fig.(\ref{fig:1}) that the event horizon  begins shifting for values of the order $\sim 10^{-4}$. For larger values of $\beta$, the location of the horizon stays the same. The non vanishing independent components of Riemann tensor for spherically symmetric spacetimes, including SBR black hole spacetime are given by \cite{wald2010general}:
\begin{eqnarray}
   R^{1}_{2 1 2} = \frac{-r f'(r)}{2} , 
   \label{15}
\end{eqnarray}
\begin{eqnarray}
    R^{1}_{0 1 0} = \frac{f(r) f''(r)}{2},
    \label{16}
\end{eqnarray}
\begin{equation}
    R^{1}_{3 1 3} = \frac{-r f'(r)\sin^{2}\theta}{2} ,
    \label{17}
\end{equation}
\begin{eqnarray}
    R^{2}_{0 2 0} = R^{3}_{0 3 0} = \frac{f(r) f'(r)}{2r},
    \label{18}
\end{eqnarray}
\begin{eqnarray}
    R^{2}_{3 2 3} = \sin^{2}\theta \bigg (1 - r^{2} f(r) \bigg ) 
    \label{19}
\end{eqnarray}
For computing the Riemann curvature tensor components in tetrad basis, we use the tetrad formalism as:
\begin{equation}
    R^{\hat{i}}_{\hat{j} \hat{k} \hat{l}} = R^{\mu}_{\alpha \beta \gamma} e^{\hat{i}}_\mu e^{\alpha}_{\hat{j}} e^{\beta}_{\hat{k}} e^{\gamma}_{\hat{l}}.
     \label{20}
\end{equation}
Following the Eq. (\ref{20}), the tidal tensor components are given by:  
\begin{eqnarray}
    R^{\hat{1}}_{\hat{0} \hat{1} \hat{0}} = \frac{f''(r)}{2},
    \label{21}
\end{eqnarray}
\begin{eqnarray}
   R^{\hat{2}}_{\hat{0} \hat{2} \hat{0}} =  R^{\hat{3}}_{\hat{0} \hat{3} \hat{0}} =  \frac{f'(r)}{2r},
   \label{22}
\end{eqnarray}

\subsection{TIDAL FORCE EQUATIONS}\label{III A}
Upon obtaining the expressions from Eq. (\ref{21}-\ref{22}), we can obtain the relative acceleration between two nearby particles as:
\begin{eqnarray}
   \frac{D^{2}\eta^{\hat{r}}}{D\tau^{2}} =  \frac{-f''(r)}{2} \eta^{\hat{r}},
   \label{23}
\end{eqnarray}
\begin{eqnarray}
    \frac{D^{2}\eta^{\hat{\theta}}}{D\tau^{2}} = -\frac{f'(r)}{2r}  \eta^{\hat{\theta}},
    \label{24}
\end{eqnarray}
\begin{eqnarray}
    \frac{D^{2}\eta^{\hat{\phi}}}{D\tau^{2}} = -\frac{f'(r)}{2r} \eta^{\hat{\phi}}.
    \label{25}
\end{eqnarray}
By substituting the values of $f(r)$ and its higher derivatives in Eq. (\ref{23}-\ref{25}) we get:

\begin{widetext}
\begin{eqnarray}
      \frac{D^{2}\eta^{\hat{r}}}{D\tau^{2}}= \dfrac{2r_\text{s}r^9-243{\cdot}2^\frac{21}{2}{\pi}^3G^3{\beta}r_\text{s}^3r+ 1067{\cdot}2^\frac{17}{2}{\pi}^3G^3{\beta}r_\text{s}^4}{2r^{12}} \eta^{\hat{r}},
    \label{26}
\end{eqnarray}
\begin{eqnarray}
      \frac{D^{2}\eta^{\hat{\theta}}}{D\tau^{2}}=  \dfrac{-5r_\text{s}r^9+243{\cdot}2^\frac{19}{2}{\pi}^3G^3{\beta}r_\text{s}^3r-485{\cdot}2^\frac{17}{2}{\pi}^3G^3{\beta}r_\text{s}^4}{10r^{12}} \eta^{\hat{\theta}},
    \label{27}
\end{eqnarray}
\begin{eqnarray}
    \frac{D^{2}\eta^{\hat{\phi}}}{D\tau^{2}} = \dfrac{-5r_\text{s}r^9+243{\cdot}2^\frac{19}{2}{\pi}^3G^3{\beta}r_\text{s}^3r-485{\cdot}2^\frac{17}{2}{\pi}^3G^3{\beta}r_\text{s}^4}{10r^{12}} \eta^{\hat{\phi}}.
    \label{28}
\end{eqnarray}
\end{widetext}
\subsection{RADIAL TIDAL FORCE}\label{III B}
\begin{figure}[h]
    \centering
    \includegraphics[width = \linewidth]{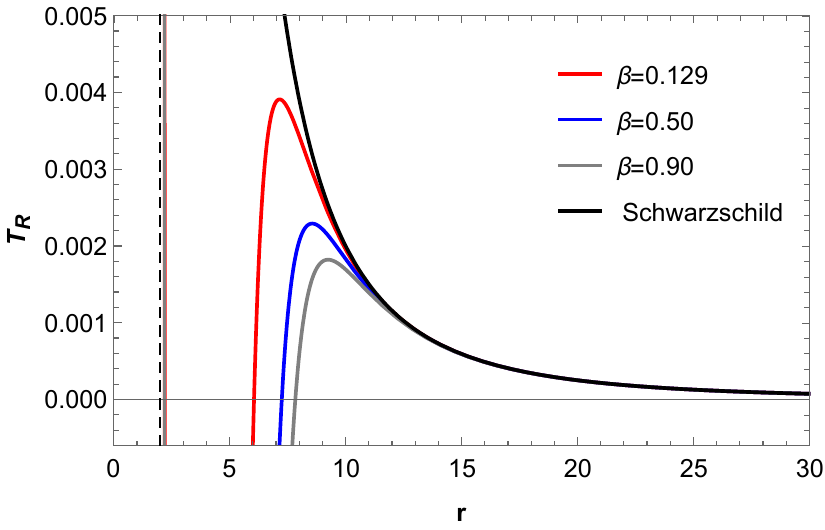}
    \caption{Radial tidal force as a function of the radial coordinate. We notice that the force vanishes at a single point in contrast to what happens with Schwarzschild spacetime. We take $M=1$. Dotted line shows the location of event horizon for Schwarzschild spacetime}
    \label{fig:2}
\end{figure}

\begin{figure}[h]
    \centering
    \includegraphics[width = \linewidth]{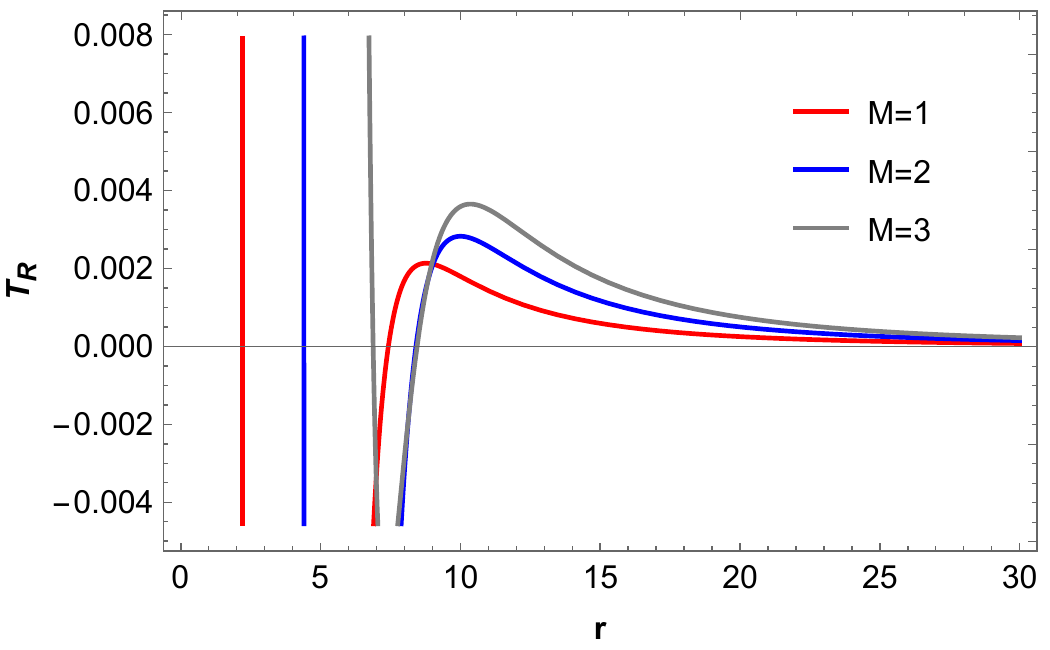}
    \caption{Radial tidal force profile for different values of M. We take $M=1$,$M=2$ and $M=3$ respectively. We use $\beta = 0.60$}
    \label{fig:3}
\end{figure}

Following the expression in Eq.(\ref{23}) and we notice that the radial tidal force vanishes at a point $R = R_0^{rtf}$ given by solving the below equation:
\begin{eqnarray}
    R_0^{rtf} = \frac{4GM}{r^{3}}-f''(\beta,r) =0.
    \label{29}
\end{eqnarray}
where the function $f(\beta,r)$ is taken for simplicity in calculation and representation, and is given by:
\begin{eqnarray}
    f(\beta,r) = \beta\left(\frac{4 \sqrt{2} \pi  G r_{s} }{r^{3}}\right)^{3}\left(\frac{108 r- 97 r_{s}}{5 r}\right).
    \label{30}
\end{eqnarray}
It can be seen from the graph in fig.(\ref{fig:2}) that as the value of $\beta$ is increased, the peak value of tidal force shifts to the right. This suggests that the maximum radial stretching happens far from the horizon as the departure from Schwarzschild geometry is significant. The peak value of radial tidal force can be calculated by equating the first derivative of the term in RHS of Eq.(\ref{23}) to zero. 
\begin{eqnarray}
   R_{peak}^{rtf} =  \frac{d}{dr} \bigg(\frac{\frac{4GM}{r^{3}} - f''(\beta,r)}{2} \bigg) =0.
   \label{31}
\end{eqnarray}
The peculiar trend that can be noticed from the graph is that the nature of force changes from radial stretching to compression for non zero values of $\beta$. This is in strict contrast to the Schwarzschild black hole case where the radial tidal force profile shows a monotonically increasing trend and becomes infinity as $r \rightarrow 0$. The transition from radial stretching to compression reverts back very close to the event horizon, after which the tidal force increases drastically till infinity. Fig(\ref{fig:3}) shows the radial force profile for higher order values of $M$. We notice that as the value of the mass of the black hole increases, the peak value for radial stretching keeps on increasing and is attained much earlier for a radially in-falling particle.

\begin{figure}
    \centering
    \includegraphics[width = \linewidth]{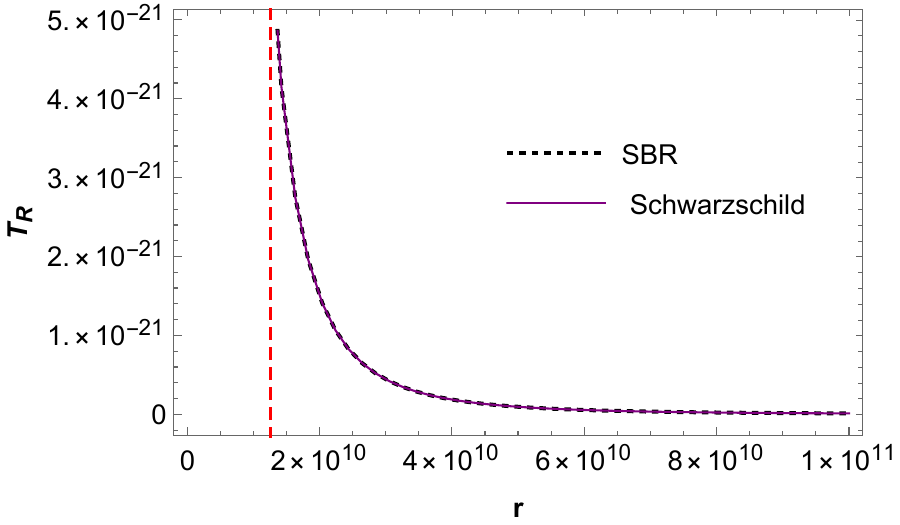}
    \caption{Radial tidal force profile for Sag A* as a function of the radial coordinate. Dotted red line shows the location of event horizon for Schwarzschild spacetime  }
    \label{fig:4}
\end{figure}

\begin{figure}
    \centering
    \includegraphics[width = \linewidth]{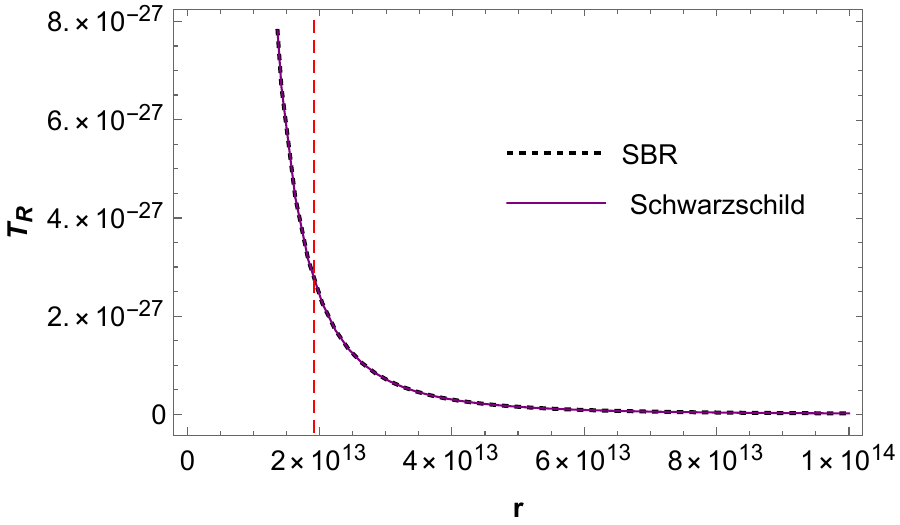}
    \caption{Radial tidal force profile for M87* as a function of the radial coordinate. Dotted red line shows the location of event horizon for Schwarzschild spacetime}
    \label{fig:5}
\end{figure}

We test the behavior of radial tidal force for both Sag A* and M87* and the results are displayed in figures \ref{fig:4} and \ref{fig:5} respectively.

For both of these cases, the force profile for SBR black hole coincides perfectly with the Schwarzschild black hole case in the range $-0.00038<\beta < 0.00089$ and  $-0.002<\beta < 0.0003$ for Sag A* and M87* respectively. This suggests that the spacetime geometry generated by Sag A* and M87* is best explained by the Schwarzschild geometry.

\subsection{ANGULAR TIDAL FORCE}\label{III C}
Just similar to the radial tidal force profile, the angular force profile also vanishes at a point $R = R_0^{atf}$. It can be computed by equating the RHS of Eq.(\ref{24}) to zero and is given by:
\begin{eqnarray}
R_0^{atf}=\frac{2GM}{r^{2}}+f'(\beta,r) =0. 
\label{32}
\end{eqnarray}
For simplicity, the term $f'(r)$ is expressed as $\frac{2M}{r^{2}} +f'(\beta,r)$. where $f(\beta,r)$ is given by the Eq.(\ref{30}). Angular tidal forces show compressive behavior as the particle falls radially inwards from infinity. The force reaches a peak value and inverts it nature to stretching. The peak value shifts to the right for increasing value of $\beta$ similar to the case of radial force profile. This suggests that maximum compression happens far from the horizon as the departure from the Schwarzschild geometry increases.

\begin{figure}[h]
    \centering
    \includegraphics[width = \linewidth]{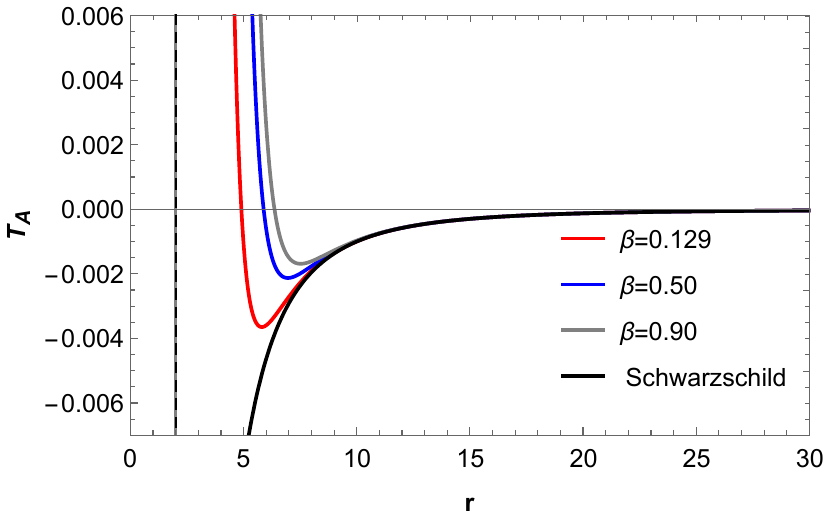}
    \caption{Angular tidal force as a function of the radial coordinate. We notice that the force vanishes at a single point in contrast to what happens with Schwarzschild spacetime. We take $M=1$. Dotted line shows the location of event horizon for Schwarzschild spacetime}
    \label{fig:6}
\end{figure}

\begin{figure}[h]
    \centering
    \includegraphics[width = \linewidth]{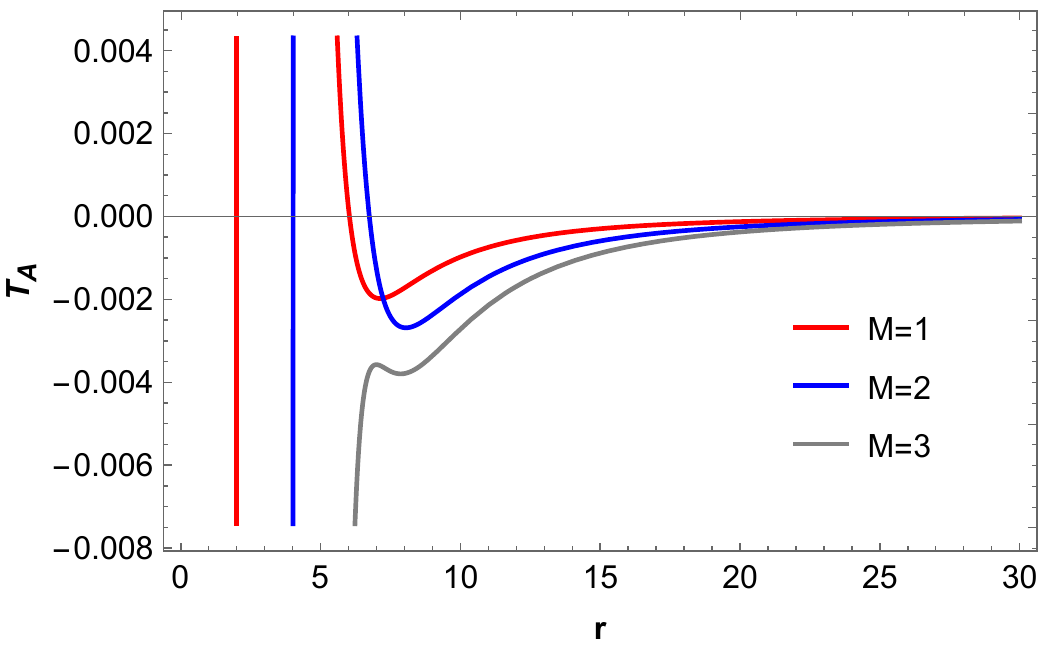}
    \caption{Angular tidal force profile for different values of M. We take $M=1$,$M=2$ and $M=3$ respectively. We use $\beta = 0.60$}
    \label{fig:7}
\end{figure}

\begin{figure}[h]
    \centering
        \includegraphics[width=\linewidth]{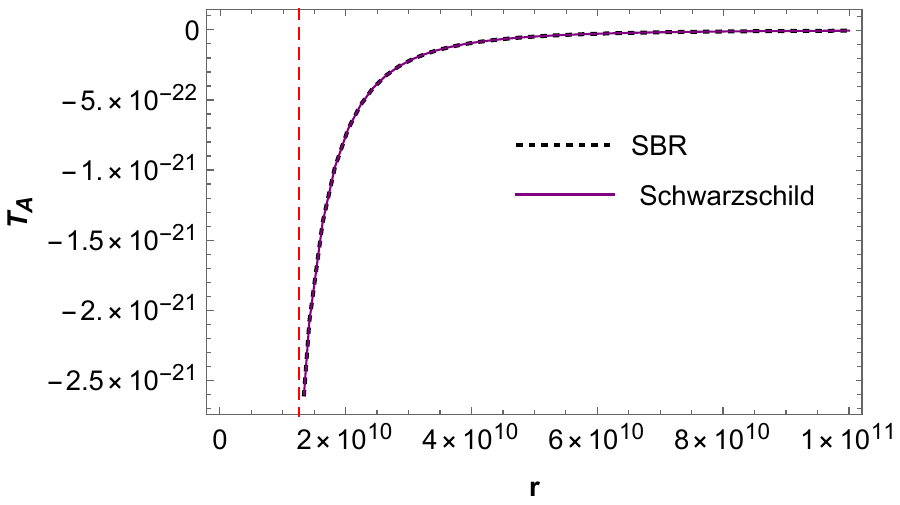}  
        \caption{Angular tidal force profile for Sag A* as a function of the radial coordinate. Dotted red line shows the location of event horizon for Schwarzschild spacetime}
        \label{fig:8}
        \end{figure}

\begin{figure}[h]  
        \includegraphics[width=\linewidth]{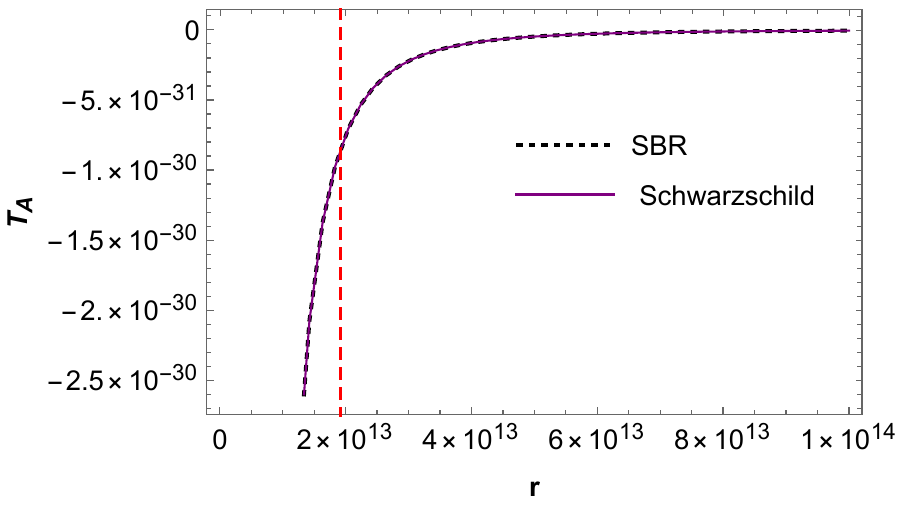}  
        \caption{Angular tidal force profile for M87* as a function of the radial coordinate. Dotted red line shows the location of event horizon for Schwarzschild spacetime}
        \label{fig:9}
    \end{figure}

Following the convention in Eq.(\ref{30}), the peak value of angular tidal force can be calculated by equating the first derivative of the expression in Eq.(\ref{24}) to zero.
\begin{eqnarray}
    R_{peak}^{atf} = \frac{d}{dr} \bigg (\frac{\frac{-2GM}{r^{2}} - f'(\beta,r)}{2r} \bigg ) =0.
    \label{33}
\end{eqnarray}
Angular force profile also shows a very different trend from the Schwarzschild geometry. In Schwarzschild black hole case, the angular tidal force becomes increasingly compressive as the radial coordinate becomes smaller only to reach infinity as $r \rightarrow 0$. But, in the case of SBR black hole, the angular tidal force switches its nature from  compressive to stretching as the radial coordinate decreases for in-falling particle. The graph in fig.(\ref{fig:6}) shows the general force profile. The compression to stretching trend again reverts back very close to the horizon, tending to infinity thereafter. Fig(\ref{fig:7}) shows the angular force profile for higher order values of $M$. We notice that as the value of the mass of the black hole increases, the peak value for angular compression keeps on increasing, indicating an increase in the tidal force on the radially in-falling particle by the black hole.
We test the behavior of angular tidal force for both Sag A* and M87* and the results are displayed in figures \ref{fig:8} and \ref{fig:9} respectively. We find that for both the cases, the SBR and Schwarzschild force profile coincides. This confirms that the spacetime geometry generated by both compact objects resembles Schwarzschild geometry.

\newpage
\section{GEODESIC DEVIATION}\label{IV}
In this section we discuss the variation of geodesic deviation vector with the radial coordinate in SBR black hole spacetime. The deviation vector measures the deformation of a body falling radially in any spacetime geometry. We can convert Eq. (\ref{23}) and (\ref{24}) in second derivatives w.r.t. $r$ by substituting $dr/d\tau$ = $-\sqrt{E^{2}-f(r)}$ which results from Eq. (\ref{9}). This gives us the following second order differential equations in $r$:
\begin{eqnarray}
  ( E^2-f(r)) \frac{D^{2}\eta^{\hat{r}}}{D r^{2}} - \frac{f'(r)}{2} \frac{D \eta^{\hat{r}}}{dr}+\frac{f''(r)}{2} \eta^{\hat{r}} = 0,
  \label{34}
\end{eqnarray}
\begin{eqnarray}
    ( E^2-f(r)) \frac{D^{2}\eta^{\hat{i}}}{D r^{2}} - \frac{f'(r)}{2} \frac{D \eta^{\hat{i}}}{dr} + \frac{f'(r)}{2r} \eta^{\hat{i}} =0.
    \label{35}
\end{eqnarray}
where $i = \{\theta,\phi\}$. The analytic solution for Eq. (\ref{34}) and (\ref{35}) as pointed out in \cite{crispino2016tidal} can be given as:
\begin{eqnarray}
  \eta^{\hat{r}} (r) = \sqrt{E^{2}-f(r)}\bigg[C_{1}+C_{2}\int \frac{dr}{(E^{2}-f(r))^{3/2}}\bigg] ,
  \label{36}
\end{eqnarray}
\begin{eqnarray}
  \eta^{\hat{i}} (r) = \bigg [C_{3} +C_{4} \int \frac{dr}{r^{2} \sqrt{(E^{2}-f(r))}} \bigg ] r.
  \label{37}
\end{eqnarray}
where $C_{1}, C_{2}, C_{3}, C_{4}$ are constants of integration. In order to find their value, we numerically solve the differential equations by imposing some initial conditions. For the purpose of this study we take the following initial conditions.
\begin{eqnarray}
    \eta^{\hat{\beta}} (b) >0 , \dot{\eta}^{\hat{\beta}} (b) = 0.
    \label{38}
\end{eqnarray}
where $\beta = \{r,\theta,\phi\}$. $\eta^{\hat{\beta}} (b)$ represents the separation between two nearby geodesics at $r=b$ in the radial and angular directions. The initial condition in Eq. (\ref{38}) represents a mass released from rest at $r=b$. In the next sections we will discuss the components of radial and the angular deviation vectors in detail.

\subsection{RADIAL COMPONENT}\label{IV A}
\begin{figure}[h]
    \centering
        \includegraphics[width=\linewidth]{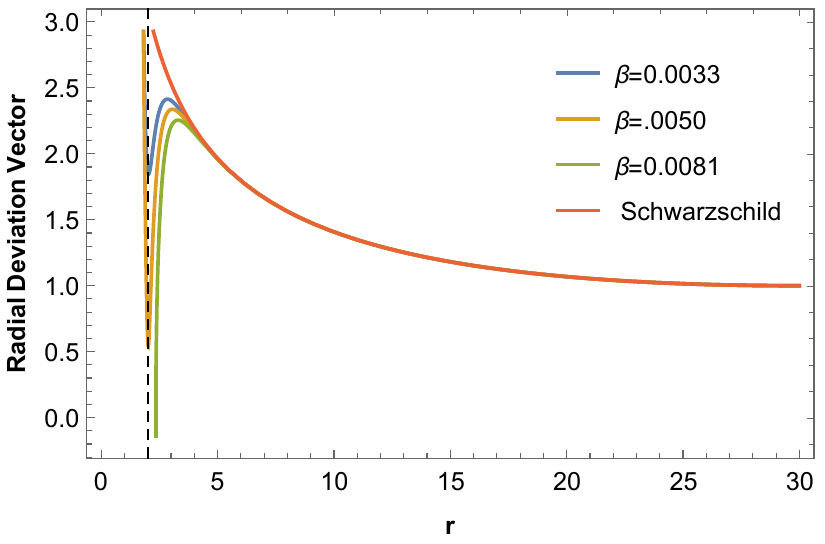}  
        \caption{Radial geodesic deviation profile for a radially in-falling particle in SBR black hole spacetime. We use M=1 to show the trend for different values of $\beta$. Dotted line shows the location of event horizon in Schwarzschild spacetime.}
        \label{fig:10}
    
\end{figure}

\begin{figure}[h]
    \centering
        \includegraphics[width=\linewidth]{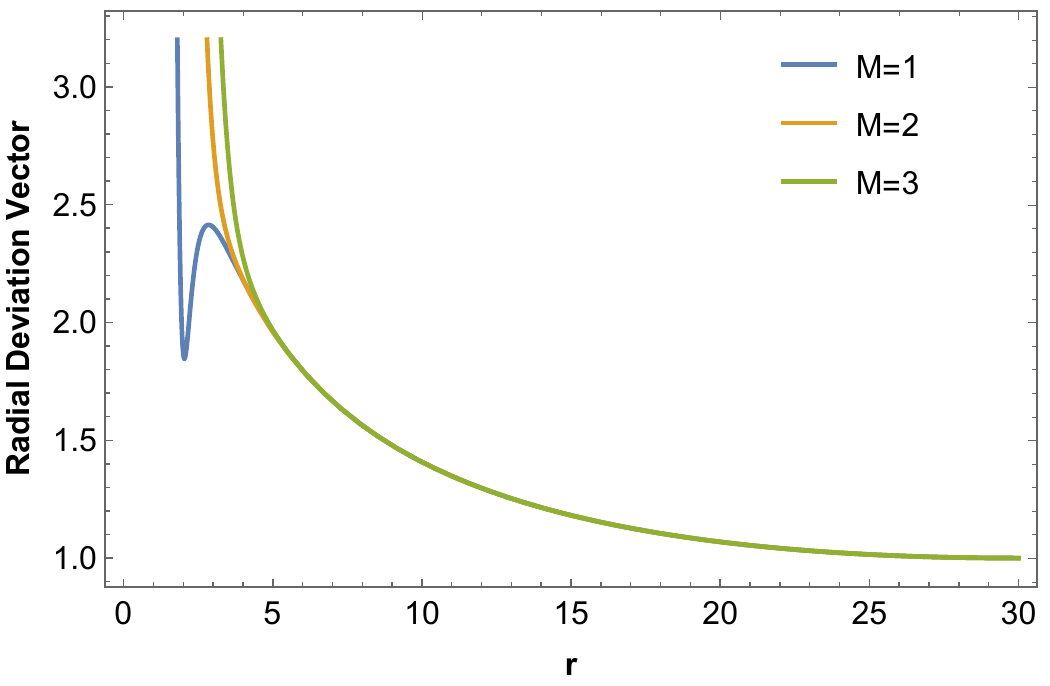}  
        \caption{Radial geodesic deviation profile for a radially in-falling particle for higher values of black hole mass M. We use $\beta = 0.0033$.}
        \label{fig:11}
    
\end{figure}

In Fig.(\ref{fig:10}) we have shown the radial component of the geodesic deviation vector after solving Eq. (\ref{36}) with initial condition. We notice that for non zero values for $\beta$, the radial component of the geodesic deviation vector reaches a finite value outside the event horizon, falls rapidly and again peaks just inside the horizon as $r \rightarrow 2M$, unlike in the Schwarzschild black hole scenario, where it crosses the event horizon and goes till infinity because of the infinite stretching radial tidal force at the singularity. This peak value attained outside the horizon keeps on decreasing as the value of $\beta$ is increased. The radial geodesic deviation trend for higher values of mass can be seen in Fig.(\ref{fig:11}). For $M=1$ case the deviation profile shows an oscillating trend. Subsequently, for higher values of $M$, the curve becomes more steep signifying a rapid increase in the separation of two nearby geodesics as the radially in-falling particle approaches the spacetime singularity. For all the cases, the geodesic deviation approaches infinity as $r \rightarrow 0$.
 
\begin{figure}
    \centering
        \includegraphics[width=\linewidth]{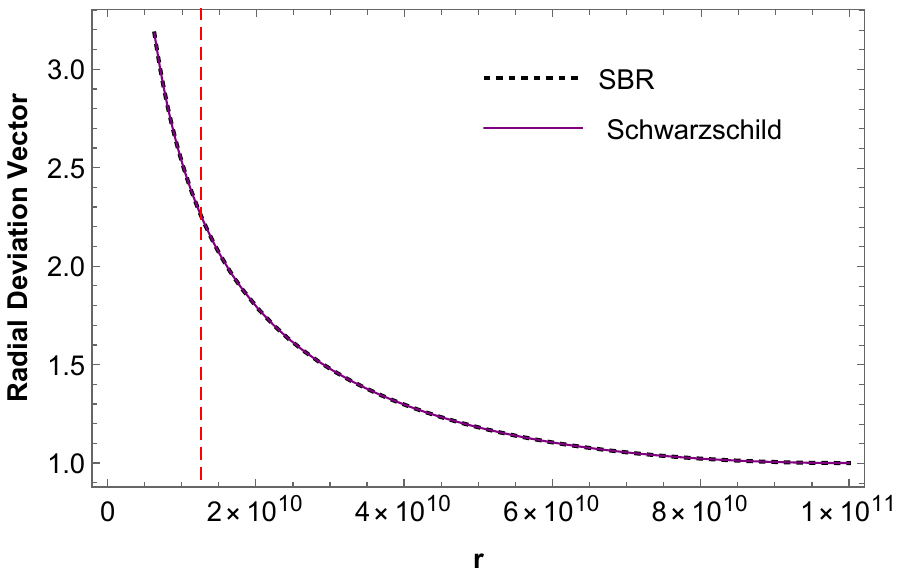}  
        \caption{Radial geodesic deviation profile for Sag A* using the constrained values of $\beta$. The dotted red line shows the location of the event horizon of Schwarzschild spacetime. }
        \label{fig:12}
    \end{figure}

\begin{figure}
    \centering
        \includegraphics[width=\linewidth]{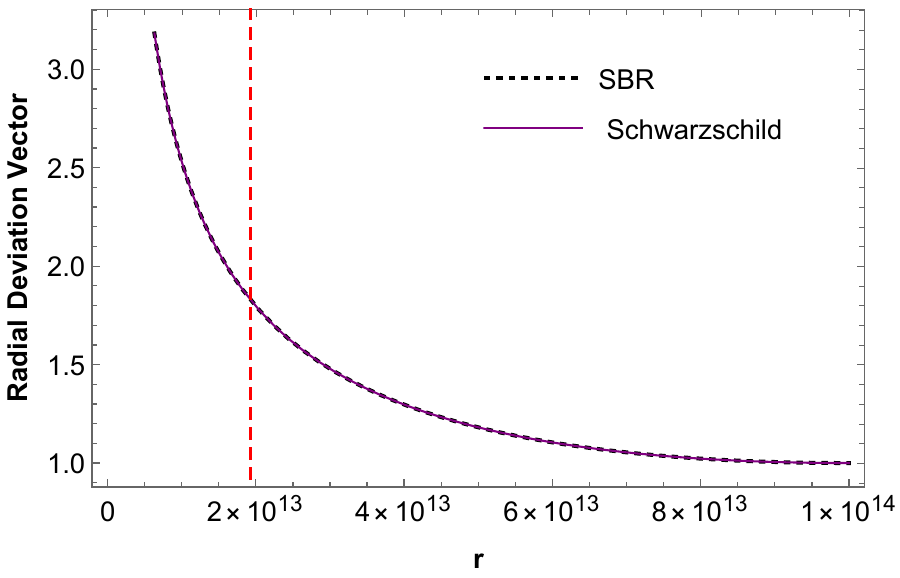}  
        \caption{Radial geodesic deviation profile for M87* using the constrained values of $\beta$. The dotted red line shows the location of the event horizon of Schwarzschild spacetime. }
        \label{fig:13}
    \end{figure}

By using the constrained value of $\beta$, we show the radial geodesic deviation trend for Sag A* and M87* in Fig (\ref{fig:12}) and Fig.(\ref{fig:13}). As seen from figures, the SBR black hole and Schwarzschild black hole lines perfectly coincide, leading to the conclusion that Schwarzschild geometry perfectly defines the geodesic deviation for a test particle falling radially in the spacetime generated by Sag A* and M87* for  $-0.00038<\beta < 0.00089$ and  $-0.002<\beta < 0.0003$ respectively.

\subsection{ANGULAR COMPONENT}\label{IV B}
\begin{figure}[h]
    \centering
        \includegraphics[width=\linewidth]{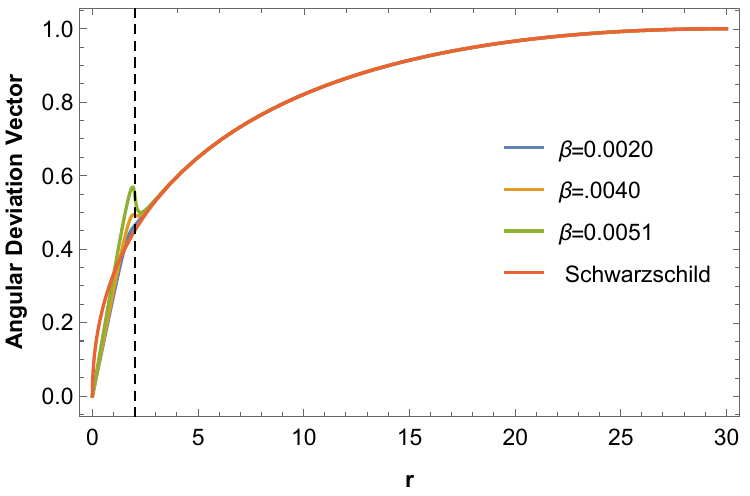}  
        \caption{Angular geodesic deviation profile for a radially in-falling particle in SBR black hole spacetime. We use M=1 to show the trend for different values of $\beta$. Dotted line shows the location of event horizon in Schwarzschild spacetime.}
        \label{fig:14}
    \end{figure}

\begin{figure}[h]
    \centering
        \includegraphics[width=\linewidth]{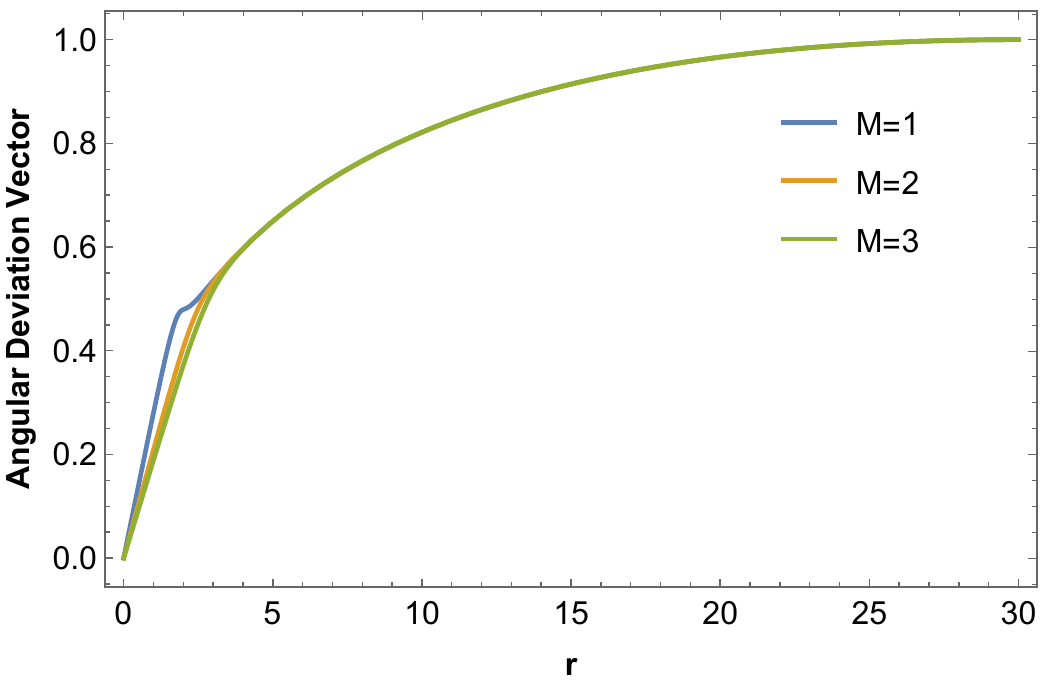}  
        \caption{Angular geodesic deviation profile for a radially in-falling particle for higher values of black hole mass M. We use $\beta = 0.0033$.}
        \label{fig:15}
    
\end{figure}

In Fig.(\ref{fig:14}) we have shown the angular component of geodesic deviation vector by constraining the Eq. (\ref{37}) with the initial condition. The results are compared with the Schwarzschild black hole. It is observed that in the SBR black hole case, the angular deviation decreases for a radially in-falling particle, shows a little spike near the event horizon and then continues to fall beyond the event horizon. This is in contrast to the Schwarzschild black hole scenario in which the deviation decreases throughout the range of the radial coordinate. The graph in Fig.(\ref{fig:14}) shows the behavior of the geodesics. The peak value is higher for increasing value of $\beta$. The graph in Fig.(\ref{fig:15}) shows the angular geodesic deviation profile for higher values of black hole mass $M$. It is noticed that for increasing values of mass, the slope of the deviation vector increases, signifying a rapid decline of the separation vector of two nearby geodesics as the in-falling particle approaches the spacetime singularity.

\begin{figure}
    \centering
        \includegraphics[width=\linewidth]{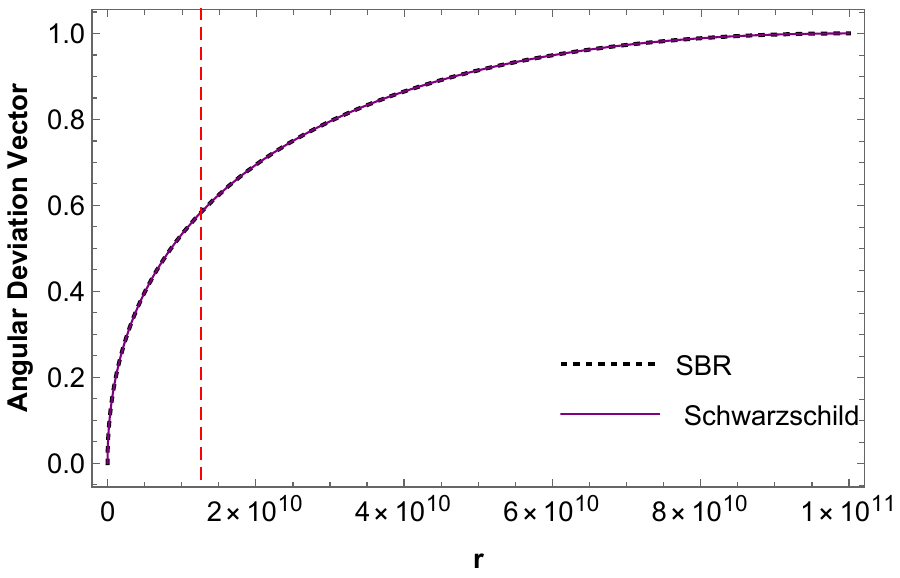}  
        \caption{Angular geodesic deviation profile for Sag A* using the constrained values of $\beta$. The dotted red line shows the location of the event horizon of Schwarzschild spacetime.}
        \label{fig:16}
    \end{figure}

\begin{figure}
    \centering
        \includegraphics[width=\linewidth]{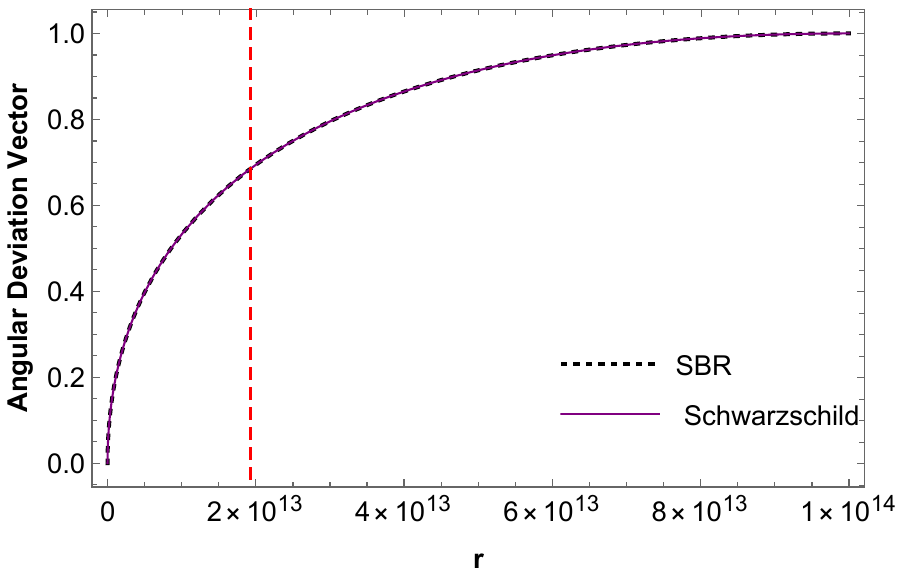}  
        \caption{Angular geodesic deviation profile for M87* using the constrained values of $\beta$. The dotted red line shows the location of the event horizon of Schwarzschild spacetime.}
        \label{fig:17}
    \end{figure}

By using the constrained value of $\beta$, we show the angular geodesic deviation trend for Sag A* and M87* in Fig (\ref{fig:16}) and Fig.(\ref{fig:17}). As seen from the graphs below, the SBR black hole and Schwarzschild black hole lines perfectly coincide, leading to the conclusion that Schwarzschild geometry perfectly defines the geodesic deviation for a test particle falling radially in the spacetime generated by Sag A* and M87* for  $-0.00038<\beta < 0.00089$ and  $-0.002<\beta < 0.0003$ respectively.

\section{BLACK HOLE SHADOW WITH \text{M87*} and \text{Sgr A*}}\label{V}

In this section we explore the shadow of the BH SBR gravity from M-theory. 
We know that for the angular radius of the BH shadow we study \cite{Perlick2022rev,Atamurotov22a}
\begin{eqnarray}\label{shadow nonrotating1}
\sin^2 \alpha_{sh}=\frac{h(r_{ph})^2}{h(r_{obs})^2},
\label{39}
\end{eqnarray}
with
\begin{eqnarray}
h(r)^2=\frac{r^2}{f(r)},\label{eq:h}
\label{40}
\end{eqnarray}
where $r_{ph}$, $r_{obs}$ are the radius of photon sphere and observer distance, respectively. $h(r)$ was early introduced in several references (for example, see \cite{Atamurotov22a}). $\alpha_{sh}$ is the angular radius of the BH shadow.

Here we combine Eqs.~(\ref{shadow nonrotating1}) and ~(\ref{eq:h}), and for a distant observer the Eq.~(\ref{shadow nonrotating1}) takes the following form

\begin{eqnarray}\label{shadow nonrotating2}
\sin^2 \alpha_{sh}=\frac{r_{ph}^2}{f(r_{ph})}\frac{f(r_{obs})}{r^2_{obs}}.
\label{41}
\end{eqnarray}

One can easily find the observable radius of BH shadow for observer at infinite distance using Eq. (\ref{shadow nonrotating2}) in the following form as\cite{Perlick2022rev}

\begin{eqnarray}\label{shadow nonrotating3}
R_{sh}&\simeq&r_{obs} \sin \alpha_{sh} \simeq \frac{r_{ph}}{\sqrt{f(r_{ph})}}.
\label{42}
\end{eqnarray}



\begin{figure}
\includegraphics[width=0.45\textwidth]{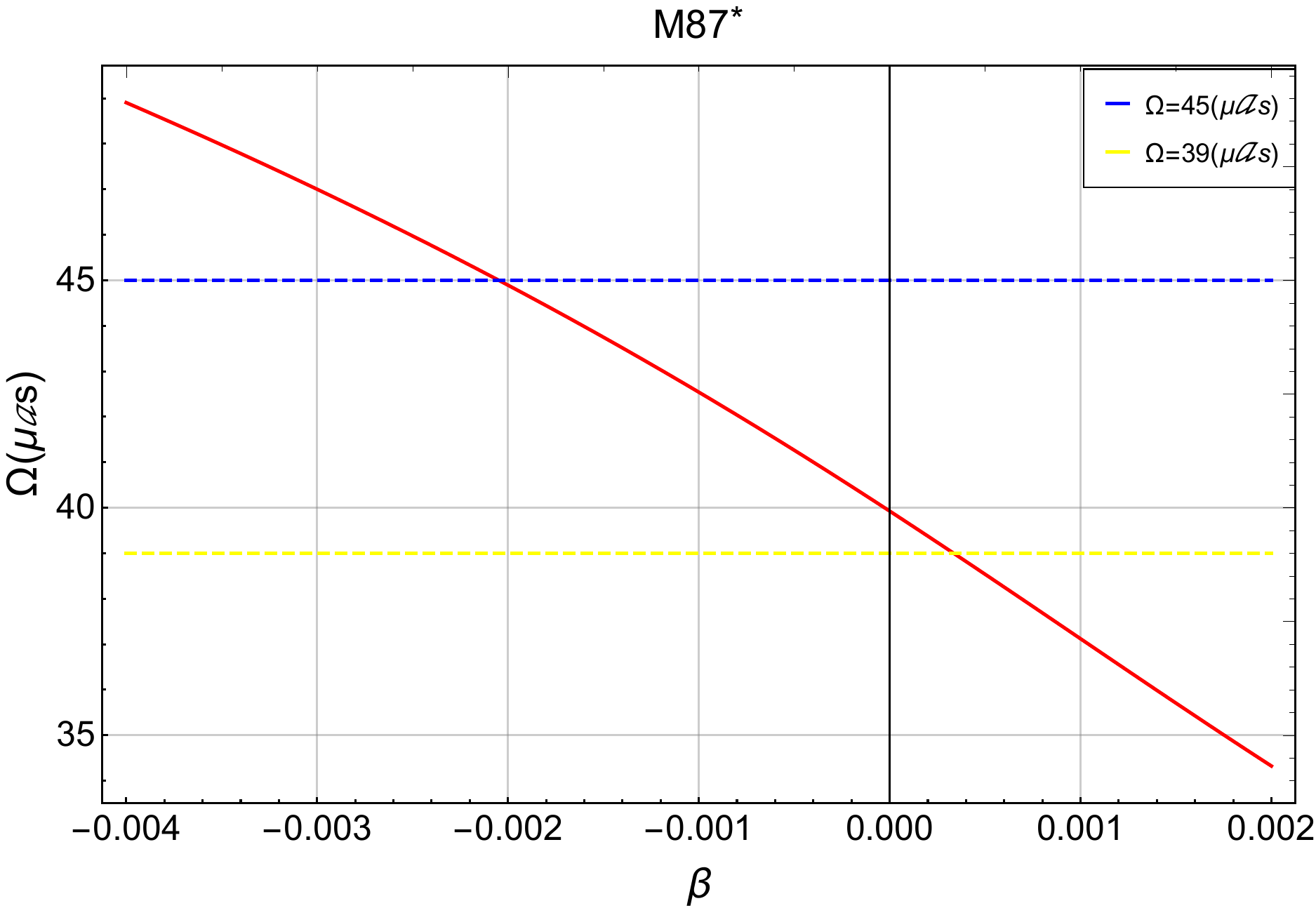}
\includegraphics[width=0.45\textwidth]{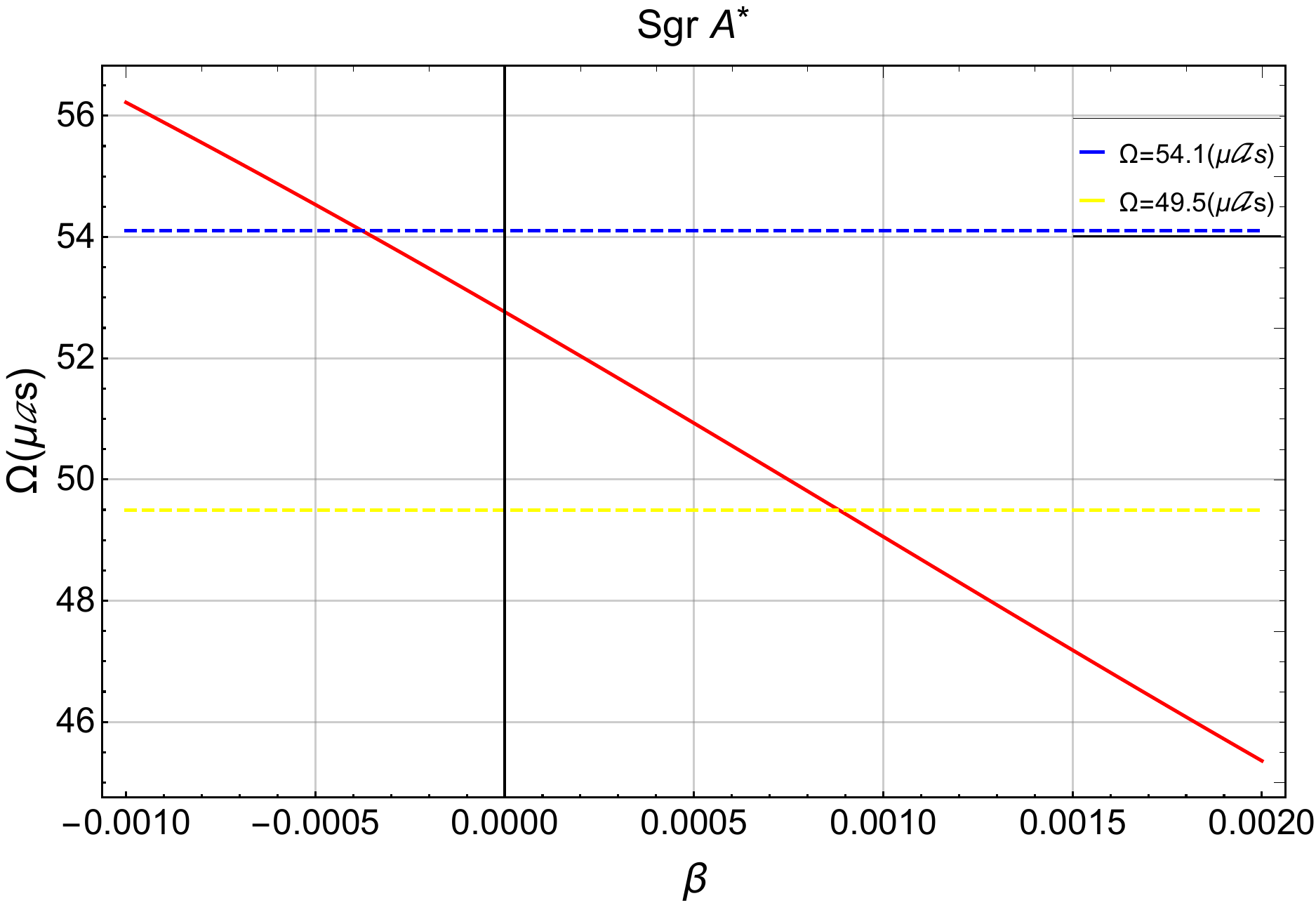}
 \caption{The angular diameter of shadow for SBR black hole as a function of parameter $\beta$ where angular diameter $\Omega=42 \pm 3 \mu as$ for M87* (upper panel); and for $\Omega = 51.8 \pm 2.3 \mu as$ for Sgr A*
(lower panel).}
 \label{fig:18}
\end{figure}

Now we consider that the supermassive BHs M87* and Sgr A* are spherically symmetric static and SBR parameter $\beta$ from M-theory. Although the observation got by the EHT collaboration does not support the assumption taken here. However, here we explore theoretically the constrain on the parameter $\beta$, from the data provided by the EHT project. To constrain this parameter $\beta$ in SBR gravity we use the observational data released by the EHT project for the BH shadows of the supermassive BHs M87* and Sgr A*. The angular diameter of the shadow, the distance from sun system and the mass of of the BH at the centre of the galaxy M87, are $\Omega_\text{M87*} = 42 \pm 3 \:\mu$as,  $D = 16.8 \pm 0.8$ Mpc and $M_\text{M87*} = (6.5 \pm 0.7)$x$10^9 \: M_\odot$, respectively \cite{2}. For the Sgr A* the data recently obtained by the EHT project is  $\Omega_\text{Sgr A*} = 51.8 \pm 2.3 \:\mu$as, $D = 8277\pm9\pm33$ pc and $M_\text{Sgr A*} = 4.297 \pm 0.013$x$10^6 \: M_\odot$ (VLTI) \cite{Akiyama2022sgr}. Using this data, we can estimate the diameter of the shadow cast by the BH, per unit mass from the following expression \cite{Bambi:2019tjh,Vagnozzi:2022moj},
\begin{equation}
    d_\text{sh} = \frac{D \Omega}{M}\,.
    \label{43}
\end{equation}
Now we can obtain the diameter of the shadow from the expression $d_\text{sh}^\text{theo} = 2R_\text{sh}$. Thus, the diameter of the BH shadow image is 
$d^\text{M87*}_\text{sh} = (11 \pm 1.5)M$ for M87* and  $d^\text{Sgr A*}_\text{sh} = (9.5 \pm 1.4)M$ for Sgr A*. From the data by the EHT collaboration, we obtain the constrain on the parameter $\beta$ for the supermassive BHs at the centre of the galaxy M87* and the Sgr A*. We present our results obtained here in the Fig.~\ref{fig:18}. In this figure,it is observed that the angular diameter of BH shadow decreases with the increasing value of parameter $\beta$. It is also observed that the angular diameter of shadow for M87* and Sgr A* BH in the context of the SBR black hole BH is smaller than the other ordinary astrophysical BH such as Schwarzschild BH. Consequently, the SBR black hole parameter $\beta$ has been constrained as $-0.00038<\beta < 0.00089$ for Sgr A* and  $-0.002 <\beta < 0.0003$  for M87* respectively.

\section{CONCLUSIONS}\label{VI}
In this work, we have discussed tidal force effects and constrained the shadow of Schwarzschild type black holes in Starobinsky-Bel-Robinson modified theory of gravity. This work has been done in an attempt to define some observational signatures that can be used to distinguish SBR gravity from Einstein's general theory of relativity. The  metric for spherically symmetric and static spacetime in this gravity differs from the Schwarzschild metric by introducing a stringy gravity parameter $\beta$ which has a positive value.\\\\
We see that both the radial and angular tidal force profiles depend on the value of $\beta$ and approach infinity as $r \rightarrow 0$. Both force profiles show a varying trend where the radial force becomes compressive and angular tidal force becomes of stretching nature as a particle radially falls towards the black hole. This is in contrast with Schwarzschild black hole case in general relativity, where the radial tidal force monotonically increases till infinity showing stretching and the angular tidal force monotonically increases till infinity and shows compression. Both tidal forces vanish at two points outside the event horizon for different values of $\beta$ giving us an opportunity to observe this phenomenon in some cases. In case of SBR gravity, the event horizon of black hole forms prior to reaching a radial distance of $r=2M$. Another important observation that can be made from the radial and angular force profiles is that the the peak value is attained farther away when compared with the Schwarzschild black hole case.\\\\ 
In addition to this, we also examined the geodesic deviation for a radially in-falling particle in the Schwarzschild like spacetime in SBR gravity. We consider the initial condition where a body is released from rest at a far away point where the spacetime is Minkowskian. It is observed that the radial deviation vector acts very similar to that of the Schwarzschild black hole case i.e. increasing as the particle approaches the event horizon, but shows an oscillating trend outside the event horizon. Opposite behaviour can be seen for angular deviation. It mimics the Schwarzschild black hole as the particle approaches the event horizon, but shows a sudden peak just around it and continues to rapidly fall to zero. Also, we have studied the tidal forces and geodesic deviation for Sag A* and M87* in SBR gravity by constraining the value of $\beta$ using shadow size from EHT observations. We can see that in the particular $\beta$ range, the tidal force and geodesic deviation profiles and both Sag A* and M87* in SBR gravity perfectly co-incide with the Schwarzschild black hole case.\\\\
Furthermore, we also calculated the angular diameter of the shadow in SBR black hole and compared it to  Schwarzschild BH. It is observed that the angular diameter of shadow for M87* and Sgr A* BH in SBR black hole is smaller than the Schwarzschild BH. Consequently, the SBR black hole parameter $\beta$ has been constrained as $-0.00038<\beta < 0.00089$ for Sgr A* and  $-0.002 <\beta < 0.0003$  for M87* respectively.It suggests that such black holes satisfy the EHT observational constraints, and it may be possible to detect the SBR black hole and distinguish it from the other astrophysical BHs in the future.
\section*{Acknowledgements}

NUM would like to thank  CSIR, Govt. of
India for providing Senior Research Fellowship (No. 08/003(0141))/2020-EMR-I). This research is partly supported by Research Grants FZ-20200929344 and F-FA-2021-510 of the Uzbekistan Ministry for Innovative Development. G. Mustafa is
very thankful to Prof. Gao Xianlong from the Department of Physics, Zhejiang Normal University, for his kind support and help during this research. Further, G. Mustafa acknowledges Grant No. ZC304022919 to support his Postdoctoral Fellowship at Zhejiang Normal University


\bibliographystyle{apsrev4-2}%
\bibliography{tidal,Lensing,mybib} 


\end{document}